\documentclass[12pt]{article}
\usepackage{csquotes}

\usepackage{geometry,amssymb,amsmath,bm}
\usepackage{marginnote}
\usepackage{doc}
\usepackage[abs]{overpic}
\usepackage{color}
\usepackage{multirow}
\usepackage{graphicx,amsmath}
\usepackage{epstopdf}

\usepackage[abs]{overpic}
\usepackage{xcolor,varwidth}

\usepackage{url}
\usepackage[export]{adjustbox}

\usepackage{array}
\newcolumntype{L}[1]{>{\raggedright\let\newline\\\arraybackslash\hspace{0pt}}m{#1}}
\newcolumntype{C}[1]{>{\centering\let\newline\\\arraybackslash\hspace{0pt}}m{#1}}
\newcolumntype{R}[1]{>{\raggedleft\let\newline\\\arraybackslash\hspace{0pt}}m{#1}}

\usepackage{graphicx,amsmath}
\usepackage{epstopdf}

\title{Multisector method for arteries: \\
The residual stresses of circumferential rings\\ with non-trivial  openings   }
\author{Taisiya Sigaeva$^a$, Michel Destrade$^b$,  Elena Di Martino$^a$\\[18pt]
\small{ $^a$ Department of Civil Engineering, 
Schulich School of Engineering,}\\[-4pt]
\small{ Libin Cardiovascular Institute, University of Calgary,}\\[-4pt]
\small{Calgary, Alberta, Canada;}\\[4pt]
\small{$^b$School of Mathematics, Statistics and Applied Mathematics,}\\[-4pt]
\small{NUI Galway, University Road Galway, Ireland.}
}
\date{}

\begin{document}

\maketitle

\begin{abstract}
The opening angle method is a popular choice in biomechanics to estimate residual stresses in arteries.
Experimentally, it means that an artery is cut into rings; then the rings are cut  axially or radially allowing them to open into  sectors; then the corresponding opening angles are measured to give residual stress levels by solving an inverse problem. However, for many tissues, for example in  pathological tissues, the ring does not open according to the theory into a neat single circular sector, but rather creates an asymmetric geometry, often with abruptly changing curvature(s). This phenomenon may be due to a number of reasons including variation in thickness, microstructure, varying mechanical properties, etc.  As a result, these samples are often eliminated from studies relying on the opening angle method, which limits progress in understanding and evaluating residual stresses in real arteries. 
With this work we propose an effective approach to deal with these non-trivial openings of rings. 
First, we digitize pictures of opened rings to split them into multiple, connected circular sectors.
Then we measure the corresponding opening angles for each sub-sector. Finally, we can determine the non-homogeneous distribution of residual stresses for individual sectors in a closed ring configuration. 
\end{abstract}

 \newpage
 
\reversemarginpar
\section{Introduction}



The determination of residual deformations is an essential piece of information to analyze and understand the mechanical behavior of soft biological tissues. 
Any incision inevitably leads to an opening; any material extraction similarly leads to a change in the geometry of both the extracted piece and the region from which it was extracted -- all revealing the presence of residual stresses. 
The kinematics of these residual deformations, together with a proper material characterization, can be used to estimate the magnitude of the residual stresses, so that they can be accounted for when modeling complicated biological processes and mimicking in-vivo loading states.

\begin{figure}
\centering
\begin{overpic}[width=0.8\linewidth]{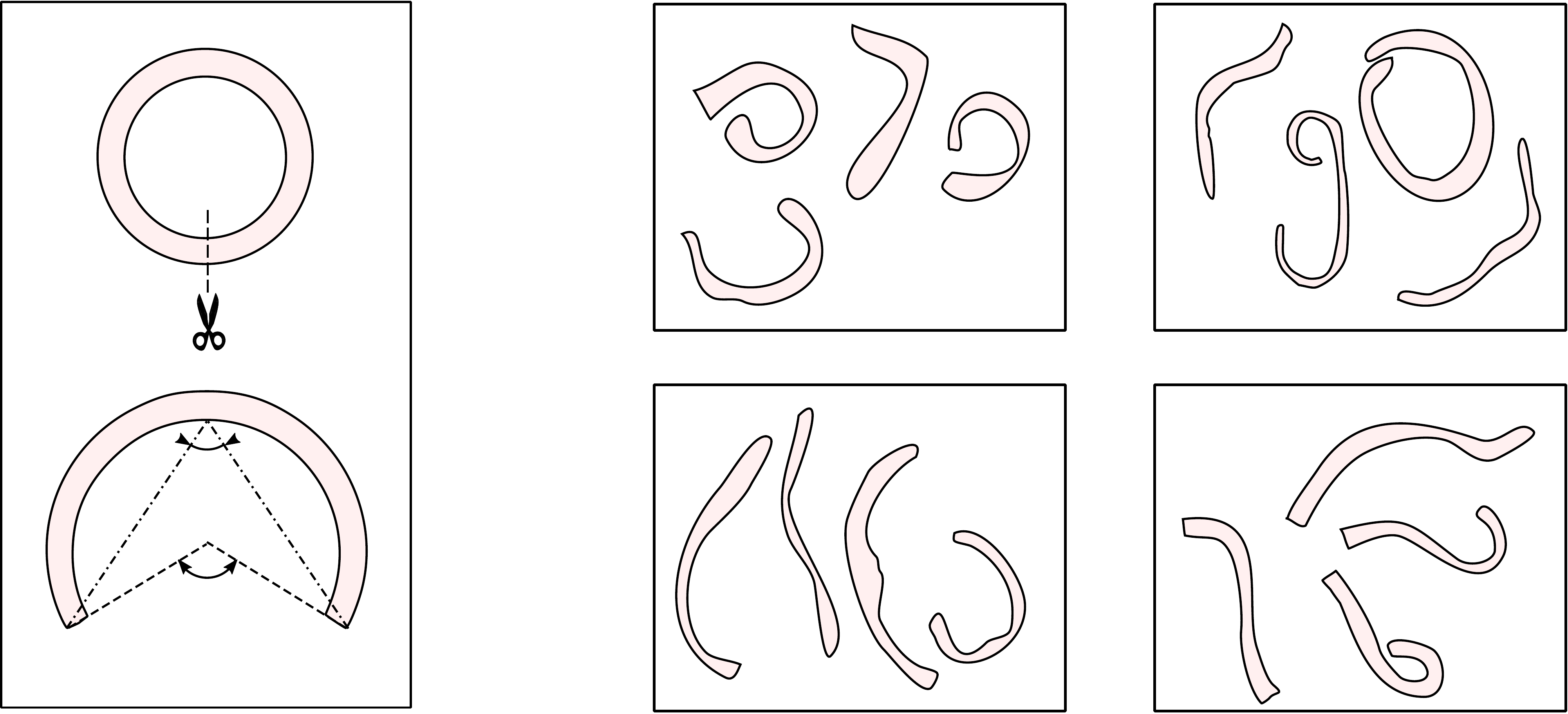}
\put(-20,147){(a)}
\put(123,147){(b)}
\put(42,52){$\varphi$}
\put(39,21){$2\alpha_r$}
\put(200,88){\cite{Fung1988,Fung1989}}
\put(255.5,88){\cite{Okamoto2002}}
\put(215.5,62){\cite{Sokolis2015}}
\put(255.5,62){\cite{Sokolis2017}}
\end{overpic}
\caption{(a) Classical opening angle method, based on the assumption of an idealised circular sector opening. (b) Sketches of real openings reported and demonstrated in the literature (references in brackets).  }
\label{fig1}
\end{figure}
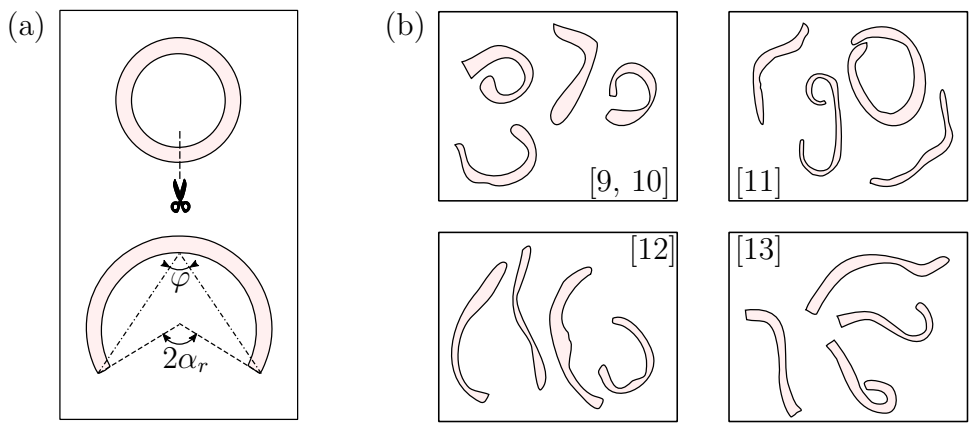

In particular  in arteries, residual stresses are known to optimize the distribution of transmural stresses due to internal pressure in such a way that the vessel wall is at its strongest compared to all its possible states \cite{Sigaeva2018}.
The conventional approach for detecting and evaluating residual stress in arteries is the \emph{opening angle method}, where thin circumferential rings from an artery are cut axially/radially and open into single sectors, revealing that the rings were under circumferentially compressive stresses, see  \cite{Vaishnav1983,Fung1984} and Figure \ref{fig1}(a). 
The opening angles are then measured and used in analytical models to estimate the magnitude and distribution of the residual stresses \cite{Fung1986}. 
Over the years, it has been become apparent that this experiment is not  capable of capturing the full residual stress distribution in all real arteries. 

First, this experiment reveals  circumferential residual stresses only, and axial stresses are neglected. 
However, an axial strip cut from the vessel can exhibit a considerable bending deformation, or undergo a  considerable change in dimensions, or can experience both phenomena at the same time. 
This simple experiment reveals that axial residual stresses exist in arteries, that they play an important mechanical role \cite{Cardamone2009} and that  it is vital to account for them. 
However, axial stresses significantly complicate the calculation of residual stresses in terms of analytical modeling, and many works prefer to neglect bending in either circumferential or axial strip, or both. 

The next limitation of the conventional opening angle approach is that it ignores that arteries are multilayered structures, and that each layer has different mechanical properties as well as different amounts and distributions of its main load-bearing components -- elastin and collagen. As a result, even though stresses that were holding axial or circumferential strips in shape while being a part of the intact blood vessel are released, the stresses to hold the artery's incompatible layers together, or interface stresses, remain. 
Hence it has been shown that the three individual aortic layers, intima, media and adventitia undergo  drastically different residual deformations upon separation \cite{Holz2007,Pena2015}, requiring involved analytical models to calculate them \cite{Holz2009}. 

Notwithstanding these limitations, the classical opening angle method introduced in the 1980s remains most popular due to its simplicity. 

In spite of decades of intense work dedicated to the study of residual stresses in soft tissues, their underlying mechanisms and their functions are still not fully understood \cite{Fung1991}. 
It is accepted that they are strongly connected to growth and remodeling, which may actually trigger  them. 
It is also established that  external factors such as hypertension or hypertrophy \cite{Fung1989a} can affect them. 
Residual stresses change with age \cite{Sokolis2017}, they are different in different arteries and even along the same aorta. Clearly, any pathological changes affecting arteries will affect residual stresses too. 
Some researchers have even hypothesized that an incorrect functioning of  residual stresses might cause pathology. 
In the aorta, for example, if the mechanism responsible for homeostatic depositing and degradation of collagen and elastin is not functioning properly, then an aneurysm is formed. 
Aneurysms may look like a localized ballooning of the aorta or may be more diffuse, but they are always associated with regional variations of properties along the circumference \cite{Sokolis2009} and localized weakening. These local weak spots may be responsible for prospective dilatation or rupture of an aneurysm, which most often causes immediate death of the patient.
That is one of the reasons why it is so important to be able to  estimate accurately residual stresses in aneurysms.

Even though there has been considerable research into the improvement of analytical models estimating residual stresses and into the overall understanding of their role, there still remains one aspect related to the classical opening angle method that has not been properly addressed. 
Many experiments show that in some tissues, more commonly in the pathological ones, some rings do not open according to the theory, into a neat single circular sector, but instead create an asymmetric geometry, often with abruptly changing curvature(s). 

Figure \ref{fig1}(b) shows sketches of the openings collected from several works: in the papers by Fung and collaborators \cite{Fung1988,Fung1989} and by Sokolis et al. \cite{Sokolis2015}, we find notably  asymmetric openings for porcine aortic rings from certain regions along the aorta; similarly in the papers by Okamoto et al. \cite{Okamoto2002} and by Sokolis et al. \cite{Sokolis2017} for human aneurysms. 
In \cite{Okamoto2002},  rings which did not remain on edge after the cut and rings in which more than half length fell over were excluded from the analysis. When a portion of the ring, but less than half, had fallen over, the opening angle was calculated using a special technique.  
Specifically, for 55 patients and 55 corresponding rings, the opening angles could not be measured in 21 cases \cite{Okamoto2002}.
In \cite{Sokolis2017}, 16\% of the total number of aortic rings were disregarded. 
These proportions may explain why there are so few studies of residual stresses in aneurysms, and also raises the question of how many samples were disregarded in other, non-aneurysmal, studies.

In this paper we propose a simple, innovative and practical method to measure residual stresses in asymmetric openings: we digitize asymmetric opened rings, split them into multiple sectors, and measure the  opening angle of each sub-sector (Section \ref{experiments}). 
Then, to determine residual stresses, we formulate the following inverse analytical problem.
Multiple undeformed sectors deform into sectors with the same curvature, and with the mathematical connection that deformed sectors should form a full ring (Section \ref{closing}). 
With this approach we can determine the non-homogeneous distribution of residual stresses for individual sectors in a closed ring configuration (Section \ref{numerical}).
One can then either interpret the non-homogeneous residual stress distribution or simply look at its average.
Analytically, this approach is not much more complicated that the classical opening angle model (see recent treatments in \cite{Sigaeva2018a,Sigaeva2018b}).

\section{Methods}



\subsection{Experiments and digital image analysis}
\label{experiments}


Here, as a case study, we  analyze the asymmetric openings of two aortic rings taken from two different aortas on separate occasions.

The first ring comes from a healthy porcine aorta (Landrace cross domestic pig, 32kg, castrated male, approximately 4 months old), particularly its upper abdominal part, denoted as ``Pig AA'' (abdominal aorta). 
The second ring comes from a human abdominal aortic aneurysm (5.7 cm aneurysm in a 55-year-old male patient), denoted ``Human AAA'' (abdominal aortic aneurysm). 
Samples were collected at the Center for Bioengineering Research and Education, Schulich School of Engineering, University of Calgary, following protocols approved by the ethical board.
They were stored at $-4^\circ$C and tested within a few hours from extraction. Both rings sliced from aortas were about 4 mm in height and cleared from connective tissues using surgical scissors.  Then they were fully immersed in room temperature phosphate buffered saline solution (PBS, pH $=7.4$) for 10 minutes and allowed to equilibrate before ``closed configuration'' imaging. 
A single cut along the anterior part of the aorta was performed using surgical scissors and the opening rings were subsequently left to rest for 10 minutes before the final ``opened configuration'' image capture. 

The digital image analysis of both closed and opened configurations was conducted using built-in and adapted functionals of Matlab (version R2018a). 
We treated differently the closed and opened configurations. 
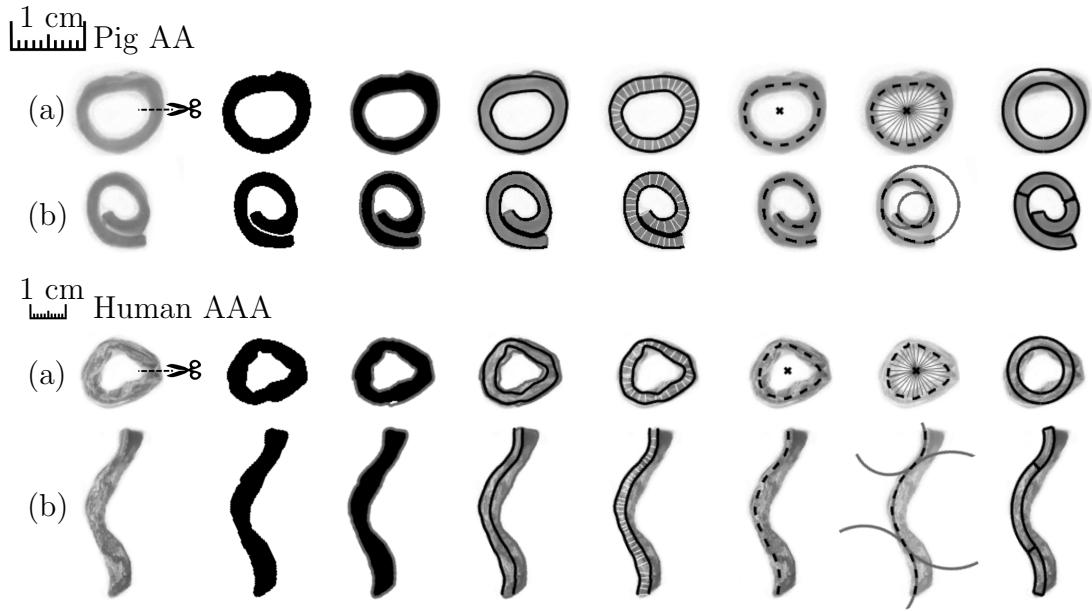
\begin{figure}
\centering
\begin{overpic}[width=1.0\linewidth]{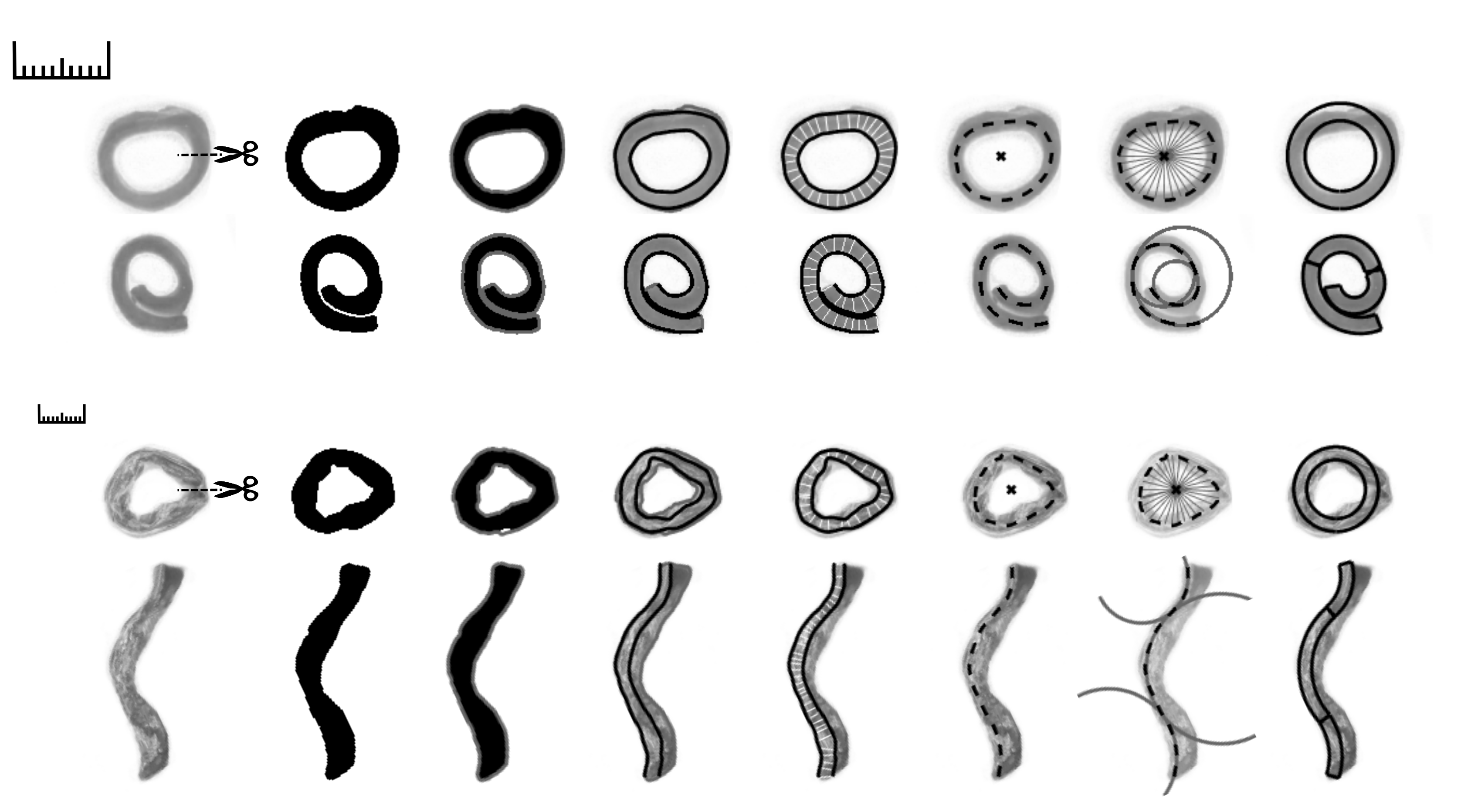}
\put(10,190){(a)}
\put(10,150){(b)}

\put(10,90){(a)}
\put(10,40){(b)}

\put(35,217){Pig AA}
\put(35,115){Human AAA}
\put(7,225){1 cm}
\put(7,122){1 cm}

\end{overpic}
\caption{Digital image analysis steps for two samples; Pig: Top row; Human: bottom row. (a): Closed configurations; (b): Opened configurations.}
\end{figure}
Closed configurations, although not perfectly circular, were considered circular. 
Here the main objective was to determine the mean aortic ring thickness and radius from images. 
These quantities fully describe the geometry of the deformed, load-free ring. 
First, the image of the ring was binarized; as the Human AAA sample had quite a bit of color variation along the thickness, different thresholds were used for Human AAA and Pig AA samples.  
Next, the Sobel-based algorithm was used to detect the edges of the rings. 
As pointed out by Holzapfel et al. \cite{Holz2007}, these edges are just approximations of real boundaries, which are impossible to determine exactly, and they have to be manually adjusted for an optimal fit. 
Here the edges were represented as natural cubic spline curves with 30 knots, and these knots were manually dragged to reach a good correspondence with the real ring boundaries. 
This manual adjustment was not necessary for the Pig AA sample but proved essential for the  Human AAA sample, as its boundaries were not clear. 
Next, the shortest distances between spline-approximated inner and outer edges were found and averaged to determine the mean thickness of the rings. 
Then the centroid of the binarized shape was found. 
Finally, the mid-line between the inner and outer spline-edges was determined and the distance from its knots to the centroid was averaged to estimate the mean radius, as depicted in Figure 2.

For the opened configurations, the initial steps were the same as for the closed configuration, but here the main goal was to determine the smallest number of individual arcs that would fit the mid-line between inner and outer spline-edges. 
To do this, the Hough circle transform was used to fit the mid-line \cite{Matlab}; the arcs of the circles intersecting at points of abrupt curvature changes were then selected to represent the mid-line. Their radii and slanted angles were determined (angles formed by rays passing through fitted circle intersections and sample's ends). It was possible to represent both Human AAA and Pig AA opened rings with just three sectors. We note that for the Human AAA there is a smooth transition between arcs and sectors, in contrast to the transitions in the Pig AA sample.





\subsection{Closing of $N$ sectors into an intact tube}
\label{closing}


Assume that we have $N$ joint sectors from different circular-cylindrical tubes. With the following plane-strain bending deformations,
\begin{equation}\label{def}
r=r^{(i)}(R),\qquad \theta=\kappa^{(i)}\Theta, \qquad z=\lambda_z^{(i)} Z,\quad (i=1...N)
\end{equation}
they form a closed intact tube occupying the region
\begin{equation}
a \le r \le b, \quad 0 \le \theta \le 2\pi, \quad 0\le z \le \ell.
\end{equation}
Here ($R, \Theta, Z$) are the coordinates of the cylindrical system aligned with the unit vectors ($\mathbf E_R, \mathbf E_\Theta, \mathbf E_Z$) in the undeformed configuration,   and  ($r, \theta, z$) and ($\mathbf e_r, \mathbf e_\theta, \mathbf e_z$) are their counterparts in the deformed configuration. Also, $\lambda_z^{(i)}$ is the axial stretch of the $i$-th sector and $\kappa^{(i)}$ reflects the \emph{change in the central angles} between the $i$-th sector's undeformed and deformed configurations.

Each sector $i=1...N$ occupies its own portion of the tube defined as
\begin{equation}
 -\alpha_d^{(i)} \le \theta^{(i)} \le \alpha_{d_i}^{(i)} , \quad\text{with}\quad\theta^{(i)}=\theta-\alpha_d^{(i)}-\sum_{k=1}^{i-1}2\alpha_d^{(k)}.
\end{equation}
Here $\theta^{(i)}$ is the conveniently chosen subsidiary circumferential coordinate used to locate the position of the deformed sectors $i=1...N$. 
When $i=1$, the sector is located at $\theta\in[0,2\alpha_d^{(1)}]$; when $i=2$, it is located at $\theta\in[2\alpha_d^{(1)};2\alpha_d^{(1)}+2\alpha_d^{(2)}]$, etc. 
Clearly, the following condition must hold,
\begin{equation}
\label{cond}
\sum_{i=1}^N\alpha_d^{(i)}=\pi.
\end{equation}

Undeformed sectors $i=1...N$ need to be either bent, unbent or everted to form a closed intact tube. The solution (\ref{def}), as the most general plane strain bending solution, is capable of capturing all these deformations via a single parameter $\kappa^{(i)}$. If we denote the central angle of the $i$-th undeformed sector as $2\alpha_r^{(i)}$, then  $\kappa^{(i)}$ can be expressed as
\begin{equation}
\kappa^{(i)} = \dfrac{\alpha_d^{(i)}}{\alpha_r^{(i)}}\in\left[-\frac{\pi}{\alpha_r^{(i)}},\frac{\pi}{\alpha_r^{(i)}}\right]\setminus\{0\},\quad \Big(\alpha_r^{(i)}\in(0,\pi],\quad \alpha_d^{(i)}\in[-\pi,\pi]\setminus\{0\}\Big).
\end{equation}
Therefore, $\kappa^{(i)} >1$ corresponds to regular plane-strain bending, $\kappa^{(i)}\in(0,1]$ corresponds to unbending, $\kappa^{(i)} <0$ corresponds to unbending beyond the configuration of a rectangular block, i.e.  eversion. 

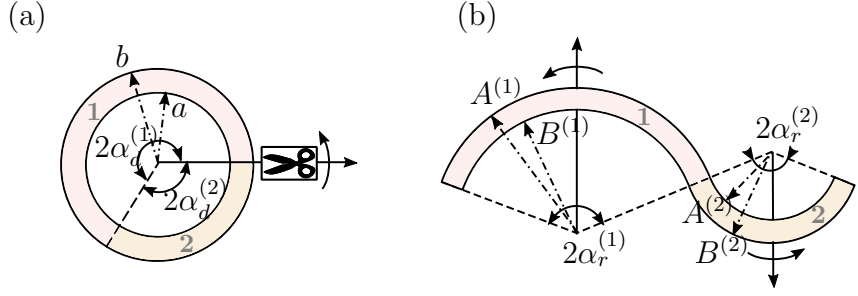
\begin{figure}
\centering
\begin{overpic}[width=0.7\linewidth]{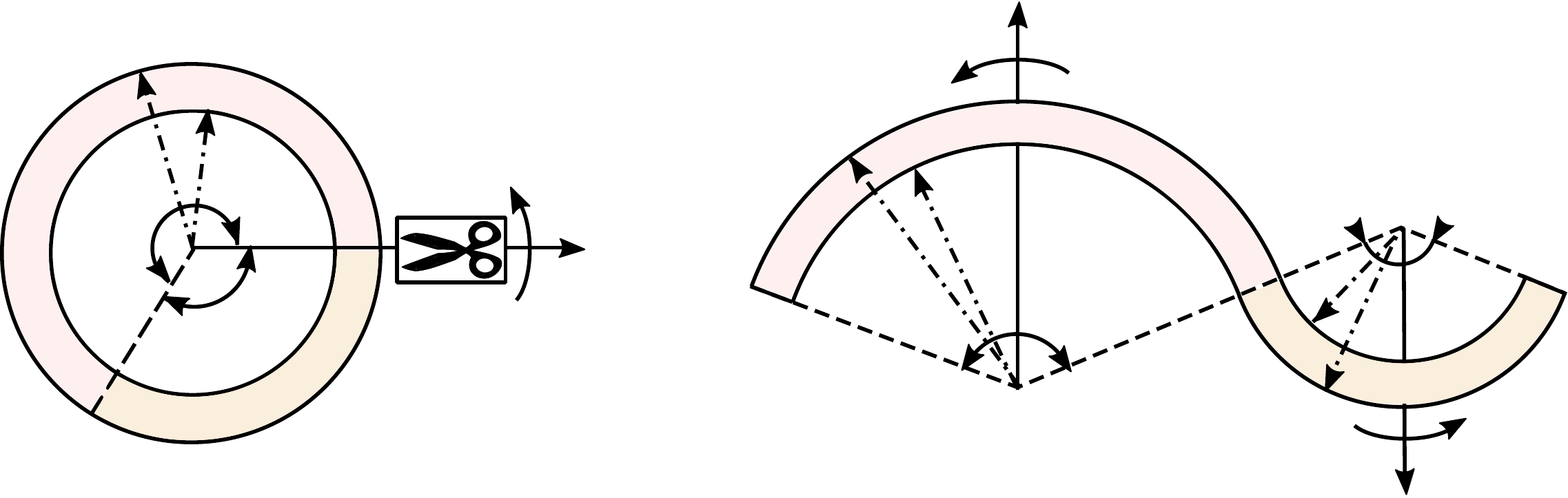}
\put(10,64){\textcolor[rgb]{0.5,0.5,0.5}{\footnotesize\text{\textbf{1}}}}
\put(45,14){\textcolor[rgb]{0.5,0.5,0.5}{\footnotesize\text{\textbf{2}}}}
\put(218,62){\textcolor[rgb]{0.5,0.5,0.5}{\footnotesize\text{\textbf{1}}}}
\put(284,25){\textcolor[rgb]{0.5,0.5,0.5}{\footnotesize\text{\textbf{2}}}}

\put(-20,100){(a)}
\put(150,100){(b)}
\put(21,84){$b$}
\put(42,65){$a$}
\put(39,30){$2\alpha_d^{(2)}$}
\put(13,50){$2\alpha_d^{(1)}$}
\put(190,10){$2\alpha_r^{(1)}$}
\put(263,57){$2\alpha_r^{(2)}$}
\put(155,70){$A^{(1)}$}
\put(180,55){$B^{(1)}$}
\put(235,25){$A^{(2)}$}
\put(240,10){$B^{(2)}$}
\end{overpic}
\caption{Deformed (a) and undeformed (b) configurations.  }
\label{configs}
\end{figure}

Assume that there are $N_1$ sectors that were bent and unbent from their undeformed states and $N_2$ sectors that were everted, so that $N_1+N_2=N$. We will use the following Sets $\mathbf{N_1}=\{i\in[1,N]:\kappa^{(i)}>0\}$ and $\mathbf{N_2}=\{i\in[1,N]:\kappa^{(i)}<0\}$ to differentiate between these sectors. 
If we define the undeformed radii as $R(a)\equiv A^{(i)}$ and $R(b)\equiv B^{(i)}$, then the sectors from Sets $\mathbf{N_1}$ and $\mathbf{N_2}$ will occupy the following regions
\begin{align}
& A^{(i)}\leq R\leq B^{(i)}, &&  -\alpha_r^{(i)} \le \Theta \le \alpha_r^{(i)}, &&  0 \leq Z \leq L^{(i)}, \qquad\text{when}\; i\in \mathbf{N_1};\\
& B^{(i)}\leq R\leq A^{(i)}, &&  \alpha_r^{(i)} \le \Theta \le -\alpha_r^{(i)}, && 0 \leq Z \leq L^{(i)},\qquad\text{when}\; i\in \mathbf{N_2}.
\end{align}

The corresponding deformation gradients $\mathbf F^{(i)}$ are
\begin{equation}\label{dgrad}
\mathbf{F}^{(i)} = \dfrac{\text d r^{(i)}(R)}{\text d R}\mathbf e_r \otimes \mathbf E_r + \dfrac{\kappa^{(i)} r}{R}\mathbf{e}_\theta\otimes\mathbf{E}_\Theta+\lambda_z^{(i)}\mathbf{e}_z\otimes\mathbf{E}_Z, \quad \lambda_z^{(i)}=\frac{l}{L^{(i)}} \quad (i=1...N).
\end{equation}
Taking the sectors to be incompressible, $\det \mathbf{F}^{(i)}=1$ must hold at all times, from which we deduce that
\begin{equation}
\label{geom}
r^{(i)} = \sqrt{\frac{R^2-(A^{(i)})^2}{\kappa^{(i)}\lambda_z^{(i)}}+a^2},
\quad 
b=\sqrt{\frac{(B^{(i)})^2-(A^{(i)})^2}{\kappa^{(i)}\lambda_z^{(i)}}+a^2} \quad (i=1...N).
\end{equation}

For sectors made of  incompressible, isotropic and hyperelastic material with strain energy  density $W=W(I_1, I_2)$, the constitutive law for the Cauchy stress tensor $\boldsymbol{\sigma}$ is
\begin{equation}
\boldsymbol{\sigma}=-p\mathbf{I}+2W_1\mathbf{B}-2W_2\mathbf{B}^{-1},
\end{equation}
where $p$ is the Lagrange multiplier introduced to ensure incompressibility, $\mathbf{I}$ is the identity tensor, $\mathbf{B}=\mathbf{F}\mathbf{F}^T $ is the left Cauchy-Green deformation tensor, $W_i={\partial W}/{\partial I_i}$ ($i=1,2$) with invariants  $I_1=\text{tr}\mathbf{B}$ and $I_2=\text{tr}\mathbf{B}^{-1}$. 
It follows that  for the  deformations (\ref{def}), the Cauchy stress tensors $\boldsymbol{\sigma}^{(i)}$ for sectors $i=1...N$ have the following structure,
\begin{equation}
\boldsymbol{\sigma}^{(i)} = \sigma_{rr}^{(i)}\mathbf e_r \otimes \mathbf e_r + \sigma_{\theta\theta}^{(i)}\mathbf{e}_\theta\otimes \mathbf{e}_\theta + \sigma_{zz}^{(i)}\mathbf{e}_z\otimes \mathbf{e}_z.
\end{equation}

As the components of the deformation gradients (\ref{dgrad}) do not depend on $\theta$ and $z$,  we readily deduce from the equilibrium equations that for each sector $p^{(i)}=p^{(i)}(r)$ only, and that  
\begin{equation}\label{equilibrium}
\dfrac{\text d \sigma_{rr}^{(i)}}{\text d r} + \frac{\sigma_{rr}^{(i)}-\sigma_{\theta\theta}^{(i)}}{r} = 0, \qquad (i=1...N),
\end{equation}
are the only non-trivial equations of equilibrium.
They are to be solved subject to following boundary conditions of traction-free inner and outer faces of the deformed sectors:
\begin{equation}
\label{bcs}
\sigma_{rr}^{(i)}(a)=\sigma_{rr}^{(i)}(b)=0, \qquad (i=1...N).
\end{equation}
Note that the normal forces on the deformed sub-sectors' end faces are all zero, and thus continuous in the closed ring.


\section{Numerical analysis}
\label{numerical}



\subsection{Solution procedure}


All the necessary geometric parameters for both the Human AAA and Pig AA samples required for the analytical modeling are summarised in Table \ref{table1}. 

We have fully determined the deformed configurations of the samples, described by the inner and outer radii $a$ and $b$. 
As for the undeformed configuration of each sample, it is described by three sectors with mid-lines radii $C^{(i)}=(A^{(i)}+B^{(i)})/2$ ($i=1,2,3$) and the corresponding referential angles $\alpha_r^{(i)}$. The undeformed thicknesses $H^{(i)}=|B^{(i)}-A^{(i)}|$ ($i=1,2,3$) of the sectors are deliberately left unfixed in the solution of the analytical problem. 
For the same reason $C^{(i)}$ is reported instead of $A^{(i)}$ and $B^{(i)}$, as these quantities can be determined once $C^{(i)}$ and $H^{(i)}$ are known. 

We note straight away that the undeformed sectors for the Pig AA sample experience bending and unbending, while the sectors for the Human AAA also undergo eversion, which is reflected in the negative  sign of $\alpha_r^{(1)}$ and $\alpha_r^{(3)}$.

\begin{table}[h!]
\centering\begin{tabular}{ |p{2cm} p{1.6cm}| p{1.6cm} |p{1.6cm}| p{1.6cm}| p{1.6cm}|  }
 \hline
 \multicolumn{2}{|c|}{Configuration}&\multicolumn{2}{c|}{Pig AA  sample}&\multicolumn{2}{c|}{Human AAA sample} \\\hline
Reference&&$\alpha_r^{(i)}$&$C^{(i)}$, mm&$\alpha_r^{(i)}$&$C^{(i)}$,mm\\
& Sector 1&$103.18^\circ$&$2.25$&$-24.61^\circ$&$15.66$\\
& Sector 2&$74.59^\circ$&$3.15$&$46.59^\circ$&$16$\\
& Sector 3&$63.59^\circ$&$4.95$&$-29.94^\circ$&$13.54$\\
 \hline
Deformed &&$a$, mm&$b$, mm&$a$, mm&$b$, mm\\
&&$3.56$&$5.33$&$6.59$&$9.47$\\
 \hline
\end{tabular}

\caption{Results from the  digital image analysis: in the reference configuration, $\alpha_r^{(i)}$ is the opening angle of the $i$th sub-sector and $C^{(i)}$ is its mid-line radius; in the deformed configuration, $a$ and $b$ are the inner and outer radii of the ring, respectively; see Figure \ref{configs}.}
\label{table1}
\end{table}

The sectors can be modelled using any material model  by prescribing the strain energy density function $W$. Ideally, this model should account for both mechanical and microstructural properties of cardiovascular tissues as well as anisotropy and nonlinearity. 
Here, for illustrative purposes, we chose the simple isotropic neo-Hookean material, with strain energy density function
\begin{equation}
\label{nh}
W^{(i)}=\frac{\mu^{(i)}}{2}(I_1^{(i)}-3),
\end{equation}
where $\mu^{(i)}$ is the shear modulus.

For each sample, there are seven equations to satisfy: the joining condition (\ref{cond}), the three equilibrium equations (\ref{equilibrium}) and the three boundary conditions (\ref{bcs}). 
Which seven unknowns should be selected to solve this problem? 

Clearly, it is quite challenging to measure experimentally the deformed angles $\alpha_d^{(i)}$ ($i=1,2,3$) formed by the sectors in the closed configuration; instead, we treat them as three variables to be determined from the analysis.  

Next, it is clear that three sectors with different referential angles $\alpha_r^{(i)}$ and mid-line radii $C^{(i)}$ do not close into sectors with the same curvature and thickness $h=b-a$.  
Here we can take one of three routes.
We can assume that sectors have different mechanical and microstructural properties, or that they have different axial deformations in closing $\lambda_z^{(i)}$, or that they have different undeformed thicknesses $H^{(i)}$ ($i=1,2,3$).
Here our choice of the strain energy density functions (\ref{nh}) makes our solution independent of the shear moduli $\mu^{(i)}$, which may thus be removed from the analysis. 
Hence, we  have to assume that the undeformed sectors opened asymmetrically either due to variations in wall thickness $H^{(i)}$  or due to differences in residual axial deformations $\lambda_z^{(i)}$ ($i=1,2,3$). 
In the first scenario we would have to assume that the $\lambda_z^{(i)}$ are the same for all sectors ($\lambda_z^{(1)}=\lambda_z^{(2)}=\lambda_z^{(3)}=\lambda_z$, say); in the second scenario, that the $H^{(i)}$ are the same for all three sectors ($H^{(1)}=H^{(2)}=H^{(3)}=H$, say). 






\subsection{Numerical results}


Tables \ref{table2} and \ref{table3} display the results from the solution procedure described in the previous section. 

When we assume that the ring opens asymmetrically due to differences in the residual axial deformations $\lambda_z^{(i)}$ ($i=1,2,3$), we find for the Pig AA sample that the undeformed sectors have to be contracted axially ($\lambda_z^{(i)}<1$) in order to close into the full ring, see Table \ref{table2}. 
The common thickness of the undeformed sectors, $H=1.71$ mm, is smaller than the thickness of the deformed ring $h=b-a=1.77$ mm. 
The undeformed sectors of the Human AAA sample, in contrast, have to be pre-stretched axially ($\lambda_z^{(i)}>1$) when closed, so that their initial thickness $H=2.96$ mm is larger than the thickness of the resulting ring $h=b-a=2.88$ mm. 

\begin{table}[h!]
\centering\begin{tabular}{ |p{1.7cm}|p{1.7cm}| p{1.7cm} |p{1.7cm}| p{1.7cm}| p{1.7cm}|p{1.7cm}|   }
 \hline 
 &\multicolumn{3}{c|}{Pig AA sample}&\multicolumn{3}{c|}{Human AAA sample} \\
 \hline
&$\alpha_d^{(i)}$&$H$, mm &$\lambda_z^{(i)}$&$\alpha_d^{(i)}$&$H$, mm &$\lambda_z^{(i)}$\\
 Sector 1&$52.97^\circ$&&$0.96$&$41.73^\circ$&&1.06\\
 Sector 2&$53.92^\circ$&$1.71 $&$0.95$&$91.38^\circ$&$2.96$&1.05\\
 Sector 3&$73.1^\circ$&&$0.94$&$46.89^\circ$&&$1.05$\\
 \hline
\end{tabular}

\caption{Results from the solution procedure when there is no variation in the wall thicknesses. Here $\alpha_d^{(i)}$ is the opening angle, $\lambda^{(i)}$ is the axial contraction found for the $i$-th sub-sector, and $H = H^{(1)}=H^{(2)}=H^{(3)}$ is the common thickness (in mm).}
\label{table2}
\end{table}

We find qualitatively similar results when we assume that the ring opens asymmetrically due to differences in the thickness $H^{(i)}$ ($i=1,2,3$) along the circumference, see Table \ref{table3}. 
This is the expected outcome as the variation between the axial pre-stretches $\lambda_z^{(i)}$ in the previous case was minor. 

Of course, we do not eliminate the possibility of the microstructure and/or mechanics causing the significant changes between undeformed and deformed sectors' thicknesses and heights; as stated earlier, this information can be easily included in the proposed model. 

\begin{figure}
\centering
\begin{overpic}[width=0.9\linewidth]{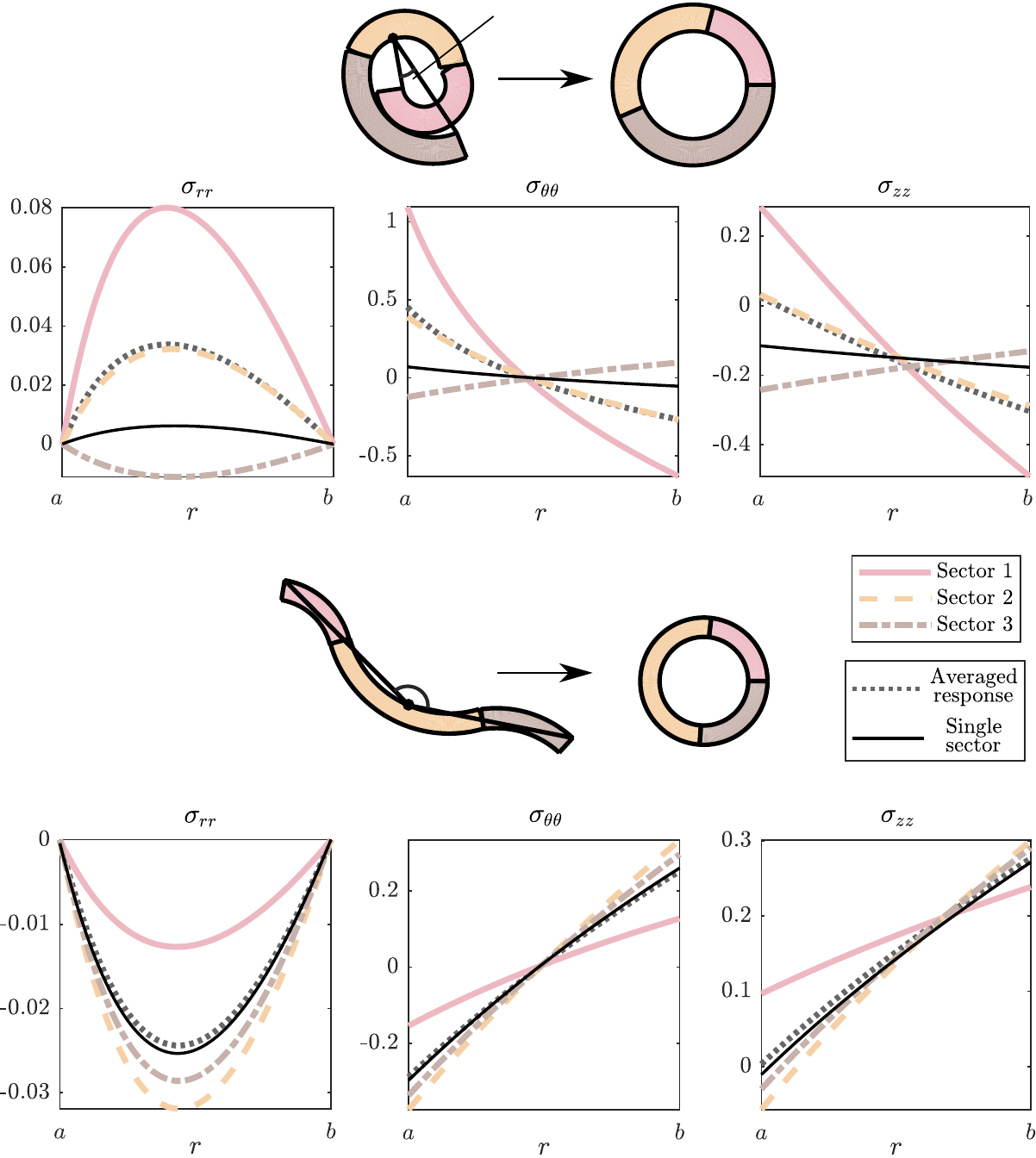}

\put(150,386.5){\textcolor[rgb]{0.5,0.5,0.5}{\footnotesize\text{\textbf{1}}}}
\put(150,421){\textcolor[rgb]{0.5,0.5,0.5}{\footnotesize\text{\textbf{2}}}}
\put(133,390){\textcolor[rgb]{0.5,0.5,0.5}{\footnotesize\text{\textbf{3}}}}

\put(275,415){\textcolor[rgb]{0.5,0.5,0.5}{\footnotesize\text{\textbf{1}}}}
\put(260,373){\textcolor[rgb]{0.5,0.5,0.5}{\footnotesize\text{\textbf{3}}}}
\put(237,415){\textcolor[rgb]{0.5,0.5,0.5}{\footnotesize\text{\textbf{2}}}}

\put(127,217){\textcolor[rgb]{0.5,0.5,0.5}{\footnotesize\text{\textbf{1}}}}
\put(145,147){\textcolor[rgb]{0.5,0.5,0.5}{\footnotesize\text{\textbf{2}}}}
\put(205,168){\textcolor[rgb]{0.5,0.5,0.5}{\footnotesize\text{\textbf{3}}}}

\put(278,203){\textcolor[rgb]{0.5,0.5,0.5}{\footnotesize\text{\textbf{1}}}}
\put(225,168){\textcolor[rgb]{0.5,0.5,0.5}{\footnotesize\text{\textbf{2}}}}
\put(283,150){\textcolor[rgb]{0.5,0.5,0.5}{\footnotesize\text{\textbf{3}}}}

\put(0,422){(a)}

\put(00,212){(b)}

\put(190,428){$\varphi$}
\put(160,180){$\varphi$}

\put(20,421){Pig AA}
\put(20,212){Human AAA}
\end{overpic}
\caption{Transmural stress components for the case where there the three sub-sectors have the same thickness ($H^{(1)}=H^{(2)}=H^{(3)}=H$).}
\label{fig4}
\end{figure}
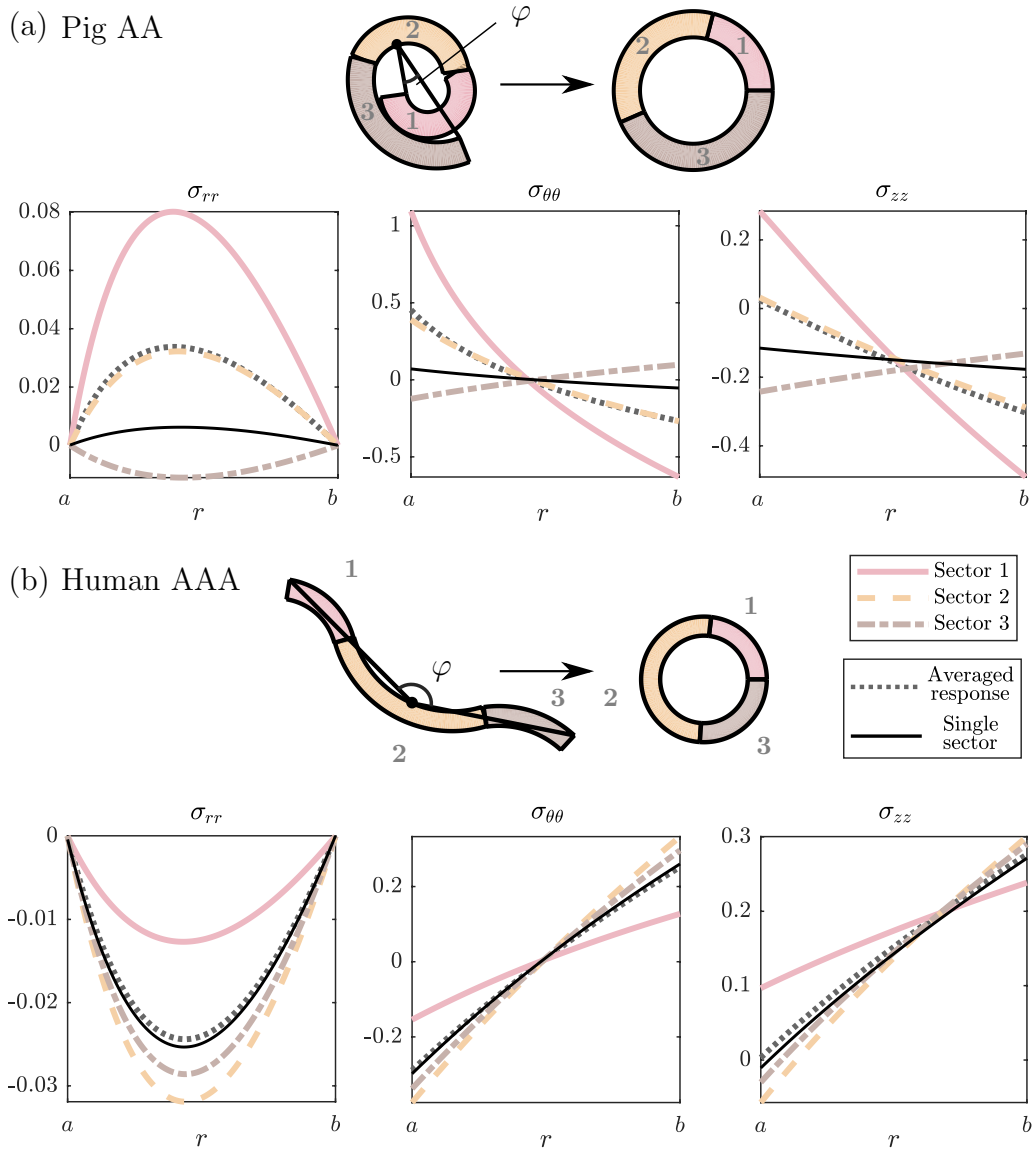

As there is no significant difference in results between the cases when asymmetric opening occurred due to variation in wall thickness $H^{(i)}$  or due to different residual axial deformations $\lambda_z^{(i)}$ ($i=1,2,3$), we depict the corresponding transmural residual stress states for all sectors of both samples for the latter case only (when $H^{(1)}=H^{(2)}=H^{(3)}=H$) in Figure \ref{fig4}.

\begin{table}[h!]
\centering\begin{tabular}{ |p{1.7cm}|p{1.7cm}| p{1.7cm} |p{1.7cm}| p{1.7cm}| p{1.7cm}|p{1.7cm}|   }
 \hline
 &\multicolumn{3}{c|}{Pig AA sample}&\multicolumn{3}{c|}{Human AAA sample} \\\hline
&$\alpha_d^{(i)}$&$H^{(i)}$, mm&$\lambda_z$&$\alpha_d^{(i)}$&$H^{(i)}$, mm&$\lambda_z$\\
 Sector 1&$53.21^\circ$&$1.70 $&&$ 41.87^\circ$&$2.95$&\\
 Sector 2&$54.01^\circ$&$1.71 $&$0.95$&$91.26^\circ$&$2.97$&$1.05$\\
 Sector 3&$72.78^\circ$&$1.72$&&$46.87^\circ$&$2.97$&\\
 \hline
\end{tabular}

\caption{Results from the solution procedure when there is no variation in the residual axial deformations.  Here $\alpha_d^{(i)}$ is the opening angle, $H^{(i)}$ is the thickness found for the $i$-th sub-sector, and $\lambda = \lambda_z^{(1)}=\lambda_z^{(2)}=\lambda_z^{(3)}$ is the common axial pre-stretch.}
\label{table3}
\end{table}

Sectors forming the opened configuration geometry of the Pig AA sample experience different types of residual stress state when closed into the ring, see Figure \ref{fig4}(a). 
The outer faces of Sectors 1 and 2 are circumferentially and axially compressed, while their inner face is under tension. 
We also observe positive radial residual stress, as is typical for unbent structures. 
Sector 3, in contrast, does not appear as stressed as the other two sectors when closed. 
It also experiences bending and, as a result, has the opposite residual stress state when part of the closed ring.


For the Human AAA sample, the sub-sectors forming its opened configuration geometry have all a very similar residual stress state once deformed into the closed ring, see Figure \ref{fig4}(b). 
All of them are unbent, and Sectors 1 and 3 are unbent to the point of eversion. 
As a result, all three sectors experience positive radial residual stress throughout, tension on their outer face, and compression on their inner face in the closed ring configuration.


\subsection{Discussion}


We now compare our treatment of  the Pig AA and Human AAA's opened rings to the one that would be expected based on the classical opening angle method. 

Okamoto et al. \cite{Okamoto2002} and by Solkolis et al. \cite{Sokolis2015} report difficulties in measuring opening angles for a large proportion of their samples. 
In these works, the opening angle is defined as the angle formed by two lines drawn from the tips of the inner circumference of the cut ring to its midpoint, see angle $\varphi$ in Figure \ref{fig1}(a). The corresponding referential angle $\alpha_r$ is related to $\varphi$ via the formula $\alpha_r=(2\pi-\varphi)/2$.
We note that rings in which more than half length fell over were excluded from these studies.  

We measured the opening angles of our samples using this technique: we found that  $\varphi=-26.38^\circ$ for the Pig AA sample, see Figure \ref{fig4}(a), and  $\varphi=146.21^\circ$ for the Human AAA, see Figure \ref{fig4}(b). 
The corresponding central angles for these openings are: $\alpha_r=193.19^\circ$ and $\alpha_r=106.9^\circ$, respectively; while $\alpha_d=180^\circ$. 
The corresponding residual stress state can be easily determined using  solution (\ref{def}) for the case $N=1$. 

To properly compare residual stresses from the classical opening angle method with the residual stresses from our multi-sector method, the following precautions need to be taken. 
First, the axial pre-stretch $\lambda_z$ in the classical opening angle method should be in line with the results of the solution for the multiple sector approach: thus, we take $\lambda_z=0.95$ for the Pig AA sample; $\lambda_z=1.05$ for the Human AAA sample, see Table \ref{table2}. 
Second, as the multi-sector approach produces different residual stresses for each sector, we compute  their average to provide a meaningful comparison with the single-sector opening angle method. 
Figure \ref{fig4} displays  both the averaged residual response from the multiple sector approach and the stresses resulting from the classical  opening angle method, calculated for a single sector. 

As can be  seen in Figure \ref{fig4}(a), the single-sector opening angle method approach clearly underestimates the residual stresses in the Pig AA ring. 
Indeed we expect that the opened ring has to experience significant circumferential tension when unfolding into a ring: here however, the opening angle method predicts a $\sigma_{\theta\theta}$ very close to zero throughout the thickness of the arterial wall.
The proposed multi-sector approach allows for a far more pronounced variation of $\sigma_{\theta\theta}$ (and indeed, of $\sigma_{rr}$ and $\sigma_{zz}$) in this sample.

Turning now to the Human AAA sample, we see that in contrast to the Pig AA, the calculated residual stresses  based either on the averaged response from the multi-sector method or on the classical opening angle method are very similar. 
However, since a big portion of the sample fell over, we may conjecture that this sample most likely would have been eliminated if the spirit of  \cite{Okamoto2002,Sokolis2015} had been adhered to. 
Our approach clearly demonstrates that the sample can be used.

The final point to touch upon in this discussion is how to use and exploit the results coming from the multi-sector approach. 
One intuitive route to follow is to use the averaged response of the stress components across the thickness, see dotted lines in Figure \ref{fig4}. 
It accounts for all opening angles in a single study, without having to eliminate any sample that does not follow the scenario of the single-sector opening angle method. 
With the averaged response we can solve the inverse problem and report the values of $\alpha_r$ and $\lambda_z$. 
For our samples, we determined the following averaged values: $\bar{\alpha_r}=251^\circ$ ($\varphi=-35.5^\circ$) and $\bar{\lambda_z}=0.95$ for the Pig AA; $\bar{\alpha_r}=108.9^\circ$ ($\varphi=142.3^\circ$) and $\bar{\lambda_z}=1.05$ for the Human AAA.

The main purpose of deriving residual stress levels is to see how they affect the in-vivo state, i.e. the transmural stresses due to internal pressure and axial tension. 
The literature shows that residual stresses can either homogenize and decrease them \cite{Destr2016}, increase them \cite{Labrosse2013}, or simply re-distribute the  stresses among the layers in a rational way by transferring larger portion of the stress to the layer designated for this purpose  \cite{Sigaeva2018}. 
The specific path taken depends on the material and microstructural properties, on the  material model and, what is more important, on the superposition between strains due to residual pre-stress and strains due to in-vivo forces in a cardiac cycle. 
Thus, even a small difference in the kinematics of residual deformations might affect the transmural distribution greatly. Here we can quantify this effect by comparing the single sector predictions in the in-vivo state (classical opening angle method) to those of each individual sector’s residual deformation, or of peak residual deformations, or of the averaged residual deformations, as in Figure 4.

Finally, another useful way to interpret these results is to turn to a regional description of the tissues. 
In some pathologically affected tissues, for example in aneurysms,  mechanical and microstructural properties vary along the circumference of the aorta. 
These variations can be matched with the asymmetric residual openings revealed by our method and used in a comprehensive finite element analysis for advanced physiological models.









\section{Conclusion}


This study proposes, demonstrates and discusses an effective way to measure residual stresses in circumferential rings that do not open in trivial and symmetric way; we call it the multi-sector method.  The asymmetric shape of the cut configuration is viewed  as being composed of several sectors, each with their individual residual stress states. 
Two examples of such openings are harvested in the lab and presented here (Pig AA and Human AAA), and their residual stresses are derived using both the proposed multi-sector approach and the classical opening angle method. 

For the Pig AA, the multi-sector approach seems to be a more physically accurate method, while for the Human AAA, it eliminates the uncertainty related to the validity of the sample. 
In practice, because each sector now  has its own residual stress state, the problem can be treated finely using finite element analysis. 
Alternatively, for analytical problems and statistical studies, either the averaged response or the peak residual stresses can  be used as metrics of overall residual stress levels.

Some limitations apply.
It is clear from the experiments that the deformed or closed configuration is not exactly circular. Accounting for a non-circular deformed configuration would require digital image splitting of the closed configuration into individual sectors, but this would complicate the analytical solution of the problem even further.

Also, the way rings open depends on where they are cut. 
But due to variation of the circumferential residual deformations along the central line of the aorta, this effect is hard to study experimentally. 
Moreover, when the opened rings are cut again, they deform even further. 
This proves that they are still under some level of residual stress, while in our model the cut configuration is split into separate stress-free sectors and, thus, the coupling between them is ignored. Using our approach for newly cut sub-sectors would make sense physically, but would introduce the uncertainty on where to make a cut. 

Moreover, in our model, for the sake of simplicity and convenience, we suggest to minimize the number of sectors describing the open configuration, which can also be seen as a limitation.

It is well known that if you cut an axial strip from the aorta, it might experience even more notable residual bending deformations than those of a circumferential ring. 
Our multi-sector approach can be adopted for the determination of residual stresses from these experiments using a block configuration as the deformed one. 

Next, our numerical results do not account for the layered structure of blood vessels, but it can be easily extended to this case. 

Finally, the residual stresses coming out of our numerical analysis do not account for the material's actual mechanical and microstructural properties, including anisotropy. 
This choice was made for the sake of simplicity, but the procedure is valid for virtually any strain energy density and more advanced constitutive models can be utilized, including properties that vary along the circumference.


The main message of this work is that  residual stresses are vital in understanding the development, normal functioning and pathology of biological tissues. 
Therefore, it is important to have a convenient tool at hand to account for the wide range of sample geometries found in practice, and for the range of tissues that do not deform in a trivial manner when residual stresses are released. 
This work proposes such a tool, as a natural extension of the classical opening angle method. 
It is capable of dealing with a good number of  samples that would have been otherwise eliminated from the analysis. 
Thanks to this versatility, more studies into the residual deformation of more challenging types of tissues might follow, particularly of aortic aneurysms, which are known to have very complicated openings as well as significant variations of properties.





\end{document}